\documentclass[floatfix,notitlepage,twocolumn]{revtex4-1}%
\usepackage{graphicx}%
\usepackage{amsmath}%
\usepackage{amsfonts}%
\usepackage{amssymb}
\usepackage{bm}
\usepackage{comment}
\usepackage{soul} 
\usepackage{multirow}
\usepackage{ulem}

\usepackage{color}
\usepackage[breaklinks]{hyperref}

\def\ve{{\varepsilon}}

\def\k{{ {\bm k} }}

\def\0{{ {\bm 0} }}

\def\expo{{ {\rm e} }}

\allowdisplaybreaks[4]

\begin{document}
\title{
Topological transitions by magnetization rotation in kagome monolayers of ferromagnetic Weyl semimetal Co-based shandite 
} 
\author{ 
Kazuki Nakazawa$^{1,2}$, \thanks{kazuki.nakazawa@riken.jp} 
Yasuyuki Kato$^1$, and 
Yukitoshi Motome$^1$ 
} 
\address{
$^1$Department of Applied Physics, The University of Tokyo, Bunkyo, Tokyo 113-8656, Japan
\\
$^2$RIKEN Center for Emergent Matter Science (CEMS), Wako, Saitama 351-0198, Japan
}

\date{\today}

\begin{abstract} 
Co-based shandite Co$_3$Sn$_2$S$_2$ is a ferromagnet hosting Weyl fermions in the layered Co kagome structure. The band topology as well as the magnetism is predicted to vary drastically in the atomically thin films depending on the thickness and surface termination, and as an extreme case, the quantum anomalous Hall state is expected in a monolayer of the Co kagome lattice. Given that the bulk Weyl gap depends on the magnetization direction, here we theoretically study how the topological nature and transport properties vary with the magnetization direction in the systems with kagome monolayer with both Sn and S surface terminations. By using {\it ab initio} calculations, we find that in the Sn-end monolayer the anomalous Hall conductivity shows successive discrete changes between different quantized values by rotating the magnetization, indicating several topological transitions between the anomalous quantum Hall insulators with different Chern numbers. Notably, when the magnetization is oriented in-plane and perpendicular to the Co-Co bond, the system exhibits a planar quantized anomalous Hall effect. We clarify that these peculiar behaviors are due to topological changes in the band structures associated with gap closing of the Weyl nodes. In contrast, the S-end monolayer shows rather continuous changes in the transport properties since the system is metallic, although the band structure contains many Weyl nodes. Our results pave the way for controlling Weyl fermions in atomically thin films of Co-based shandite, where the topological nature associated with the Weyl nodes appears more clearly than the bulk. 
\end{abstract}

\maketitle

\section{Introduction}
\label{sec:Intro}

The Weyl fermion was predicted as a massless elementary particle, which has not yet been found in nature. In condensed matter physics, however, Weyl fermions can appear as quasiparticles in particular crossing points of the electronic band structure, called Weyl nodes~\cite{AMV}. The Weyl nodes always appear in pairs and are regarded as emergent monopoles and antimonopoles in momentum space, which can be characterized by chirality $+1$ and $-1$, respectively~\cite{NN}. Their topological properties result in the so-called Fermi arcs on particular surfaces of the system as a consequence of the bulk-edge correspondence~\cite{Murakami}, and unconventional electromagnetic phenomena, such as the anomalous Hall effect~\cite{Murakami,Burkov,Suzuki}, the anomalous Nernst effect~\cite{Sakai}, and the negative magnetoresistance~\cite{Burkov,XHuang,Zhang}. A pair of Weyl nodes is generated from a Dirac node by lifting the spin degeneracy, as found in several materials with spatial-inversion symmetry breaking~\cite{Xu,Souma,WFFBD,Soluyanov,Feng} or time-reversal symmetry breaking~\cite{Suzuki,Sakai,WTVS, XWWDF,KF}. 

\begin{figure}
\includegraphics[width=90mm]{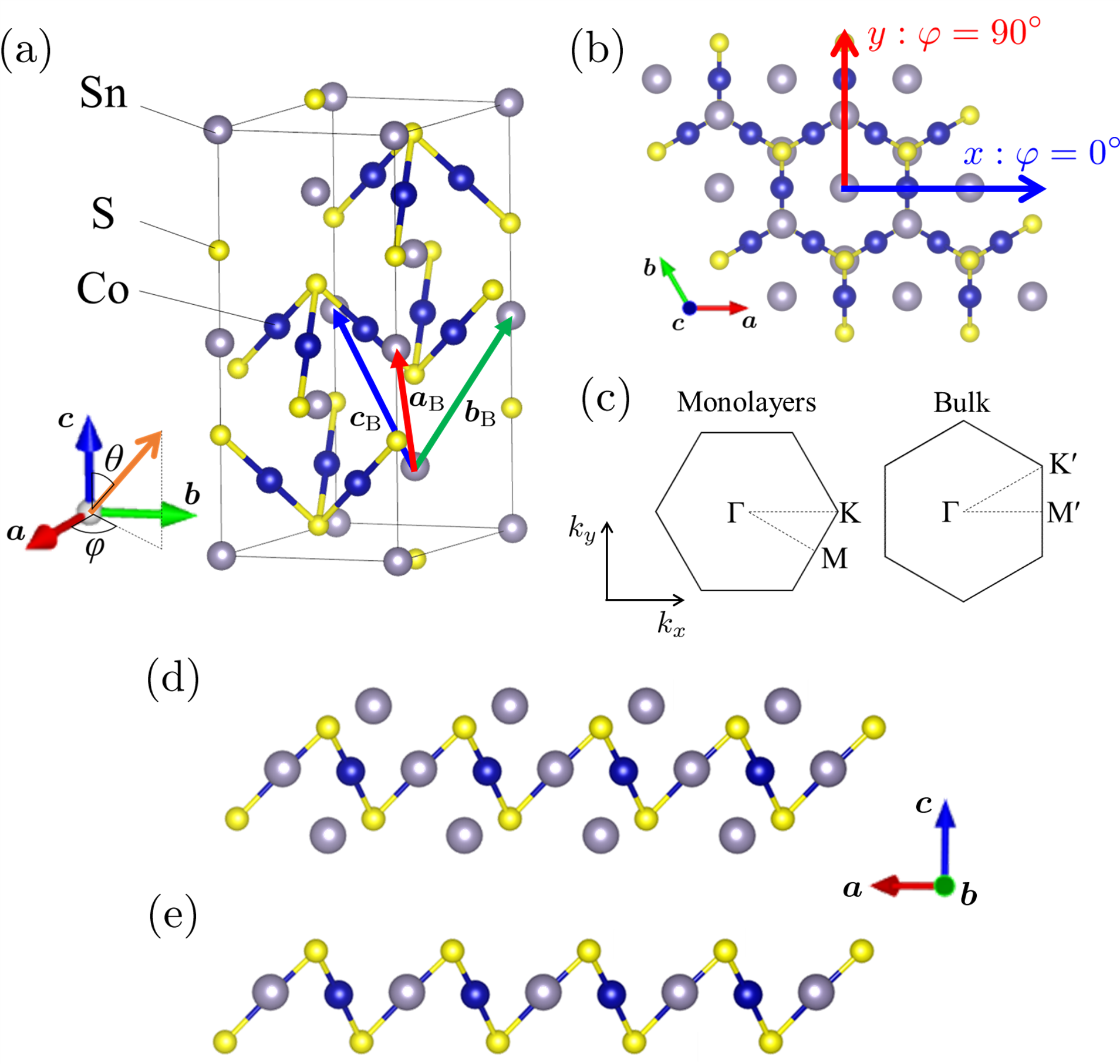}
\caption{(a) Lattice structure of the bulk Co$_3$Sn$_2$S$_2$. The black box represents the conventional unit cell containing three primitive unit cells. The inset shows the definitions of the polar and azimuthal angles, $\theta$ and $\varphi$, respectively, which specify the direction of magnetization, with respect to the lattice vectors $\bm a$, $\bm b$, and $\bm c$. ${\bm a}_{\rm B}$, ${\bm b}_{\rm B}$, and ${\bm c}_{\rm B}$ are the primitive unit vectors for the bulk system. (b) Top view of the Co kagome plane. The directions of $\varphi = 0^\circ$ and $90^\circ$ are indicated; the former (latter) is parallel (perpendicular) to a Co-Co bond. (c) Two-dimensional Brillouin zone for monolayers (left). The cut of the three-dimensional Brillouin zone for the bulk at $k_z = 0$ is also shown (right). [(d) and (e)] Side view of the (d) Sn-end monolayer and  (e) S-end monolayer. The crystal structures are visualized by VESTA~\cite{MI}.}
\label{fig:Lattice}
\end{figure}

The Co-based shandite Co$_3$Sn$_2$S$_2$, which has a stacked kagome planes composed of Co ions [Figs.~\ref{fig:Lattice}(a) and \ref{fig:Lattice}(b)], has attracted attention as a candidate for ferromagnetic Weyl semimetal with Weyl nodes associated with spontaneous breaking of time-reversal symmetry~\cite{Wang,Liu}. Indeed, this material exhibits giant anomalous Hall and Nernst effects, large negative magnetoresistance~\cite{Wang,Liu,Guin,Yang,Yanagi}, and surface Fermi arcs~\cite{DFLiu}. Subsequently, it was proposed that a thin film with Co kagome monolayer of this material realizes a Chern insulator with the quantum anomalous Hall effect~\cite{Muechler}, which has stimulated the fabrication of atomically thin films~\cite{Fujiwara,Shiogai,Ikeda1,Ikeda2,Fujiwara2}. Furthermore, it was theoretically revealed that the electronic and magnetic states, band topology, and transport properties vary significantly depending on the number of kagome layers and the surface termination in thin films~\cite{NKM}. 

In the bulk case, Co$_3$Sn$_2$S$_2$ exhibits the ferromagnetic moment perpendicular to the kagome plane~\cite{Wang,Liu}. A theoretical study based on an effective model showed that the Weyl nodes form nodal rings when the magnetization is oriented in-plane~\cite{ON}. Furthermore, a previous {\it ab initio} study demonstrated that the positions of the Weyl nodes in the energy-momentum space can be tuned by controlling the direction of the magnetic moment~\cite{Ghimire}. The magnetization angle dependence of the anomalous Hall effect as well as the spin Hall effect was also predicted for the effective model~\cite{OKN}. These results indicate that the band topology in this bulk material is sensitive to the magnetization direction. Given that the thin films of this material show a variety of the electronic and magnetic properties as mentioned above, it is interesting to clarify how these properties vary with the magnetization direction in the thin film case. It would be important not only for further understanding the relation between the magnetism and the band topology but also for further stimulating the experiments on thin film fabrication. 

In this paper, we study the electronic band structure and transport properties based on the {\it ab initio} calculations for the thin films with kagome monolayer of the Co-based shandite while changing the direction of the ferromagnetic moment. We consider two types of monolayers with Sn and S surface terminations, following the previous studies~\cite{Muechler,NKM}. We find that the anomalous Hall conductivity $\sigma_{\rm H}$ as well as the anomalous Nernst conductivity $\alpha_{\rm N}$ exhibits significantly different behaviors from the bulk system for the magnetization rotation. In particular, in the monolayer with Sn termination, we find that $\sigma_{\rm H}$ shows several discrete changes between different quantized values, suggesting topological transitions between the quantum anomalous Hall insulators with different Chern numbers. Interestingly, when the magnetization is oriented in-plane and perpendicular to the Co-Co bond, $\sigma_{\rm H}$ is nonzero at a quantized value, indicating a planar quantized anomalous Hall effect. In contrast, in the monolayer with S termination, $\sigma_{\rm H}$ shows more moderate changes without showing quantization. We explain these contrasting behaviors depending on the tilted direction of the magnetization and the surface termination by the analyses of the band structure and the Berry phase as well as the symmetry argument. 

This paper is organized as follows. After describing the details of {\it ab initio} calculations and the computations of transport properties in Sec.~\ref{sec:methods}, we show the magnetization angle dependences of $\sigma_{\rm H}$ and $\alpha_{\rm N}$ in Sec.~\ref{sec:transport_fs}. To understand the results, we discuss the electronic band structure and the Berry curvature in Sec.~\ref{sec:band_top}, and also present the symmetry argument in Sec.~\ref{sec:sym}. Finally, Sec.~\ref{sec:summary} is devoted to the summary. 

\begin{figure*}
\includegraphics[width=180mm]{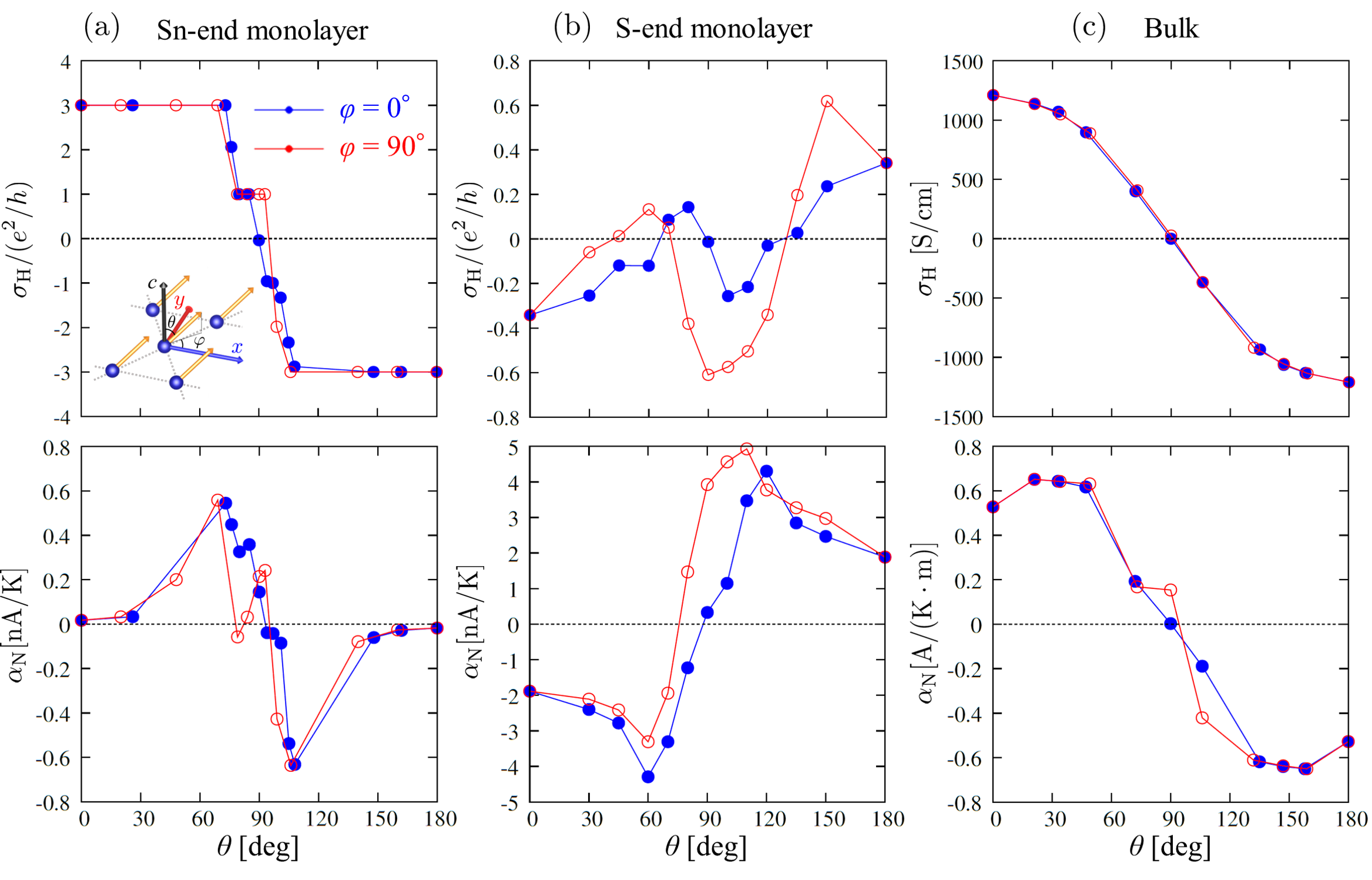}
\vspace*{-5mm}
 \caption{$\theta$ dependences of the anomalous Hall conductivity $\sigma_{\rm H}$ and the anomalous Nernst conductivity $\alpha_{\rm N}$ for (a) the Sn-end monolayer, (b) the S-end monolayer, and (c) the bulk of the Co-based shandite. $\sigma_{\rm H}$ is calculated at zero temperature, while $\alpha_{\rm N}$ is at $5~{\rm meV}$. The blue and red points represents the data for $\varphi = 0^\circ$ and $\varphi = 90^\circ$, respectively. The lines are guide for the eye. The orange arrows in the inset of the upper panel of (a) denote the Co magnetic moments.}
\label{fig:theta_dep}
\end{figure*}

\section{Method}
\label{sec:methods}

The optimized lattice structures and electronic properties are computed using the OpenMX code~\cite{OpenMX,Ozaki} within the framework of the density functional theory. The Perdew-Burke-Ernzerhof (PBE) generalized gradient approximation (GGA)~\cite{PBE} is employed for the exchange-correlation functional. We consider two types of thin films each including a Co kagome monolayer with different surface terminations: One has Sn atoms at both surfaces [Fig.~\ref{fig:Lattice}(d)], while the other has S atoms [Fig.~\ref{fig:Lattice}(e)]. We call the former and latter the Sn-end and S-end monolayer, respectively. For these monolayers, we employ the set of unit vectors $\bm a$ and $\bm b$ indicated in the inset of Fig.~\ref{fig:Lattice}(b), while we use the primitive unit vectors ${\bm a}_{\rm B}$, ${\bm b}_{\rm B}$, and ${\bm c}_{\rm B}$ in Fig.~\ref{fig:Lattice}(a) for the bulk calculations. Following the previous study~\cite{NKM}, we perform the structural optimization for the monolayers, starting from the lattice structures taken from the bulk and assuming a spin-polarized state within the nonrelativistic calculations. After the optimization, we symmetrize the structure to retain the original $p\bar{3}m1$ symmetry. In the bulk calculations, we use the experimental lattice structures~\cite{Li}. 

To study the electronic states while changing the directions of the ferromagnetic moment, we adopt the fully-relativistic calculation with a constraint on the spin orientation~\cite{OpenMX}. The magnitude of the energy constraint is taken as $5$~meV. We vary the moment direction by changing the polar angle $\theta$ from $0^\circ$ to $180^\circ$ for the azimuth angle $\varphi=0^\circ$ and $\varphi = 90^\circ$; see the inset of Fig.~\ref{fig:Lattice}(a) and Fig.~\ref{fig:Lattice}(b). Note that the cases of $\theta=0^\circ$ and $180^\circ$ correspond to the out-of-plane moments, and that the case of $\theta=90^\circ$ for $\varphi=0^\circ$ ($90^\circ$) corresponds to the in-plane moment parallel (perpendicular) to a Co-Co bond in the kagome plane. Note also that the resultant moments are roughly oriented in the direction of the constraint, but not perfectly. Such slight tilting leads to small deviations from the results expected from the symmetry, as will be discussed in Sec.~\ref{sec:sym}. We take the in-plane direction with $\varphi=0^\circ$ ($90^\circ$) as the $x$ ($y$) axis. 

To investigate the band topology and the transport properties, we construct tight-binding models from the maximally-localized Wannier functions~\cite{MV,SMV}, which are obtained by OpenMX~\cite{WOT} to reproduce the electronic band structure for each tilting angle of the ferromagnetic moment. Following the previous study~\cite{NKM}, we employ Co $3d$, Sn $5s$ and $5p$, and S $3p$ orbitals. The Berry curvature of the $n$th band at the momentum $\k$ is calculated by 
\begin{align} 
\Omega_{n} (\k) = -\sum_{m \neq n} \frac{2{\rm Im}  \left[ J_{x,nm} J_{y,mn} \right]}{[E_n(\k)-E_m (\k)]^2}, \label{eq:Berry}
\end{align}
where $m$ is the band index, $E_n (\k)$ is the eigenvalue of the tight-binding Hamiltonian $H(\k)$, and ${\bm J}_{nm}=(J_{x,nm}, J_{y,nm})$ is the current operator defined by
\begin{align}
{\bm J}_{nm} = \left\langle n \k \left\vert \nabla_\k H(\k) \right\vert m \k \right\rangle, 
\end{align}
with the eigenvector of the Hamiltonian, $\vert n \k \rangle$. The Chern number of the $n$th band is calculated by 
\begin{align}
C_n = \sum_{\k \in {\rm BZ}} \Omega_{n} (\k),
\end{align}
where the sum is taken for the momentum $\k$ within the first Brillouin zone [Fig.~\ref{fig:Lattice}(c)]. Then, based on the Kubo formula, we compute the anomalous Hall conductivity $\sigma_{\rm H}$ and the anomalous Nernst conductivity $\alpha_{\rm N}$ as 
\begin{align}
\sigma_{\rm H} &= -\frac{e^2}{\hbar V_d} \sum_{{\bm k} \in {\rm BZ}} \sum_{n} f(E_n (\k)) \Omega_{n} (\k), \label{eq:AHE} \\
\alpha_{\rm N} &= \frac{e k_{\rm B}}{\hbar V_d} \sum_{{\bm k} \in {\rm BZ}} \sum_{n} s(E_n (\k)) \Omega_{n} (\k), \label{eq:ANE}
\end{align}
respectively, where $e$ is the elementary charge, $k_{\rm B}$ is the Boltzmann constant, $\hbar$ is Dirac's constant, and $V_d$ is a $d$-dimensional volume of the system; $f(\ve) = ( \expo^{(\ve - \mu)/k_{\rm B} T} + 1 )^{-1}$ is the Fermi distribution function, where $T$ and $\mu$ are the temperature and the chemical potential, respectively, and $s(\ve)$ is the entropy density given by $s(\ve) = -f(\ve) \ln f(\ve) - [1-f(\ve)]\ln [1-f(\ve)]$. In the $\k$ summations, the number of the $\bm k$-mesh in the first Brillouin zone is taken as $2000 \times 2000$ for monolayers and $480 \times 480 \times 480$ for bulk. 

\section{Magnetization angle dependence of transport properties}
\label{sec:transport_fs}

\begin{figure*}
\includegraphics[width=180mm]{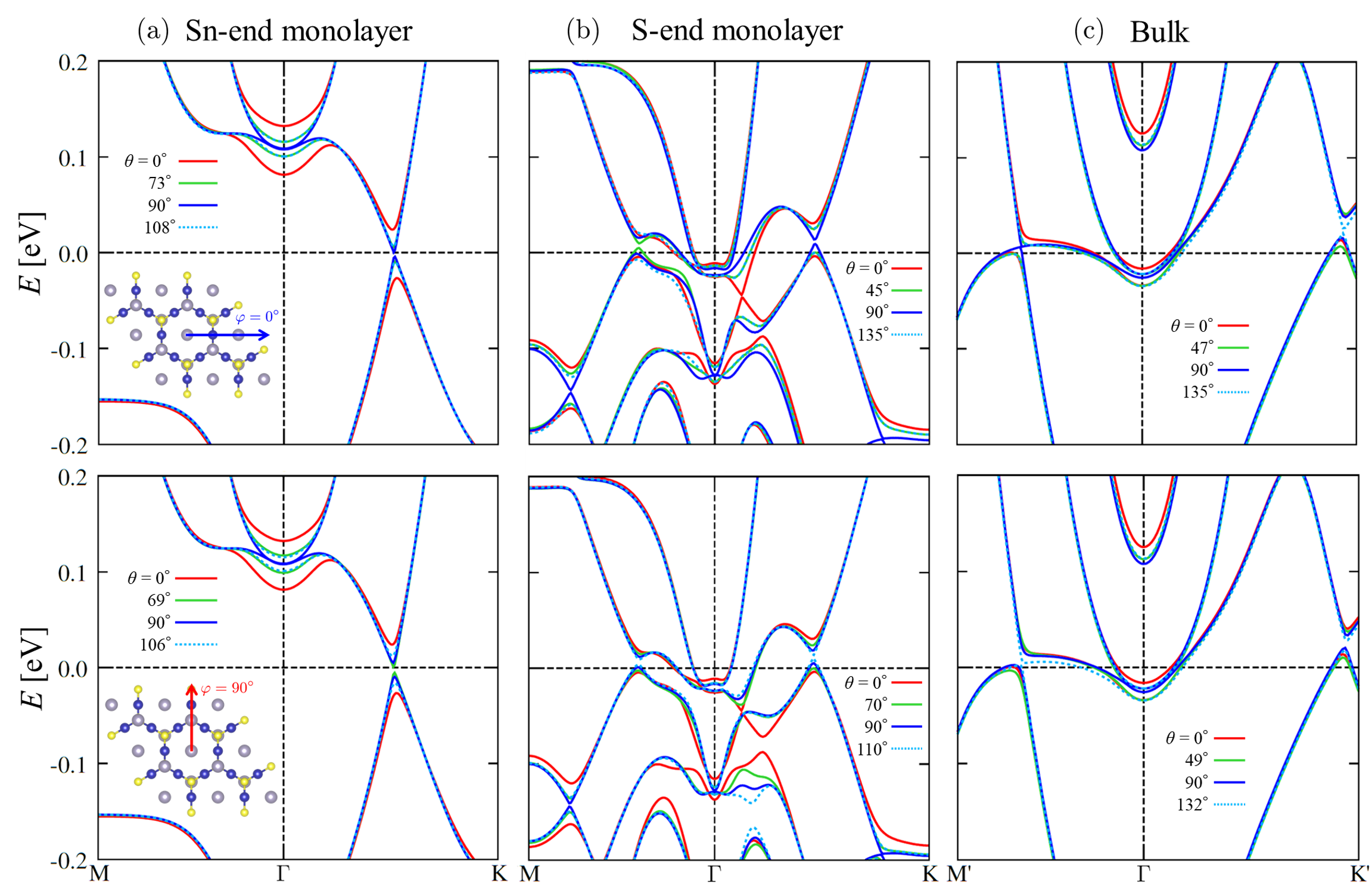}
\caption{Electronic band structures near the Fermi level ($E=0$) of (a) the Sn-end monolayer, (b) the S-end monolayer, and (c) the bulk for selected values of the tilting angle $\theta$. The upper and lower panels show the results for $\varphi=0^\circ$ and $\varphi=90^\circ$ cases, respectively.}
\label{fig:band}
\end{figure*}

In this section, we discuss the angular dependences of $\sigma_{\rm H}$ and $\alpha_{\rm N}$ while tilting the ferromagnetic moment in the $\varphi=0^\circ$ and $\varphi=90^\circ$ planes. Figure~\ref{fig:theta_dep} displays the results for the Sn-end and S-end monolayers as well as the bulk.

Let us first focus on the Sn-end monolayer [Fig.~\ref{fig:theta_dep}(a)]. In both cases of $\varphi=0^\circ$ and $90^\circ$, $\sigma_{\rm H}$ exhibits discrete stepwise changes for $\theta$. For $\varphi = 0^\circ$, $\sigma_{\rm H}$ changes between integer values in unit of $e^2/h$: $+3$, $+1$, $-1$, and $-3$. This behavior indicates that the system undergoes successive topological transitions between the quantum anomalous Hall insulating states with nonzero Chern numbers $C=-3$, $-1$, $+1$, and $+3$. We can confirm this by the plateaulike features in the chemical potential dependence of $\sigma_{\rm H}$; see Appendix~\ref{sec:transport}. In this case, $\sigma_{\rm H}$ as well as $\alpha_{\rm N}$ shows almost antisymmetric dependence on $\theta$ with respect to $\theta = 90^\circ$, indicating that the topologically trivial state with $C=0$ is realized for the in-plane magnetization with $\theta=90^\circ$. In contrast, in the case of $\varphi=90^\circ$, both $\sigma_{\rm H}$ and $\alpha_{\rm N}$ are no longer antisymmetric. In particular, $\sigma_{\rm H}$ shows stepwise changes from $+3$ to $+1$ and to $-3$ in unit of $e^2/h$, without taking $-1$, indicating that the system undergoes different sequences of topological transitions from the case of $\varphi=0^\circ$. Strikingly, $\sigma_{\rm H}$ is quantized at $+1$ even for the in-plane magnetization with $\theta=90^\circ$, suggesting a planar quantized anomalous Hall effect~\cite{LHL,Ren,ZRHZQ,ZLiu}. The contrasting behaviors between $\varphi=0^\circ$ and $90^\circ$ will be discussed from the symmetry point of view in Sec.~\ref{sec:sym}. 

\begin{figure*}
\includegraphics[width=180mm]{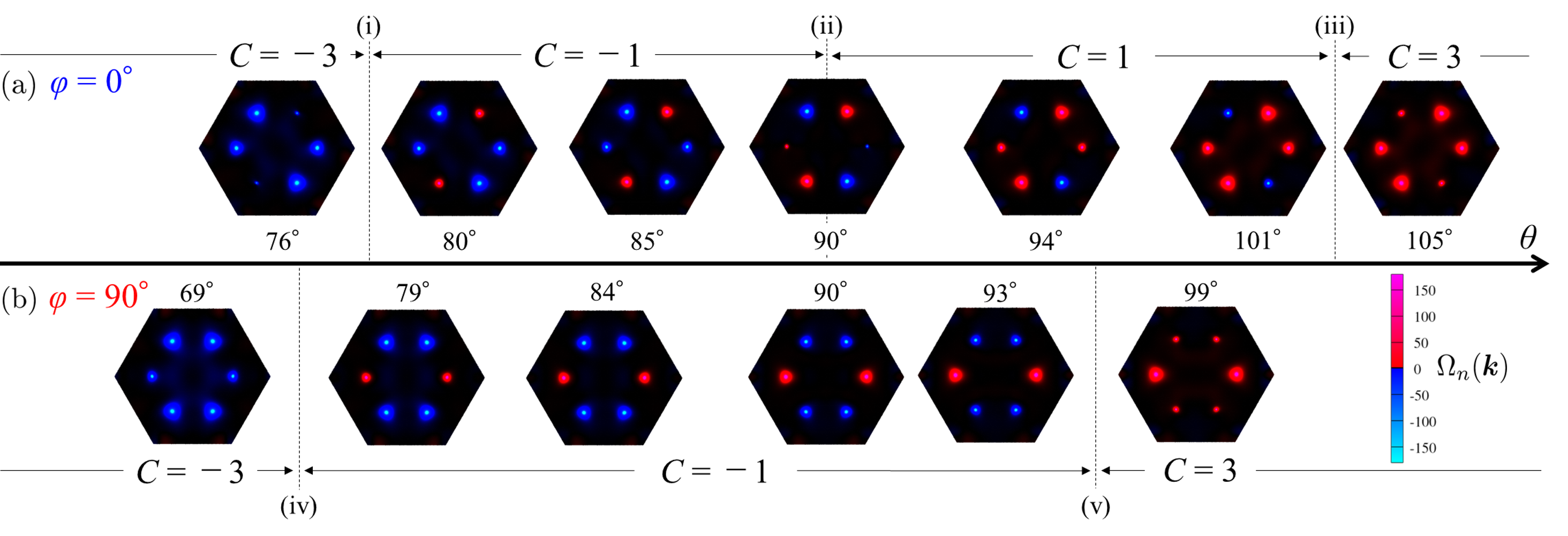}
\caption{$\theta$ dependences of the Berry curvature $\Omega_n (\k)$ in the first Brillouin zone of the band just below the Fermi level for the Sn-end monolayer: (a) $\varphi=0^\circ$ and (b) $\varphi=90^\circ$. The changes of the Chern number $C$ are also indicated. (i)-(v) represent the topological transitions associated with the changes of $C$. The band index $n$ for the Chern number is dropped.} 
\label{fig:Berry_Sn_mono}
\end{figure*}

Next, we discuss the results for the S-end monolayer [Fig.~\ref{fig:theta_dep}(b)]. In this case, $\sigma_{\rm H}$ is no longer quantized at any nonzero integer values and shows rather continuous changes with $\theta$ for both $\varphi=0^\circ$ and $90^\circ$. This is because the system is always metallic, as will be shown in Sec.~\ref{sec:band_top}. The complicated $\theta$ dependences with sign changes are ascribed to the changes of the band structure with several Weyl nodes. We note that the $\theta$ dependences of $\sigma_{\rm H}$ and $\alpha_{\rm N}$ are again almost antisymmetric with respect to $\theta=90^\circ$ for $\varphi=0^\circ$, while they are not for $\varphi=90^\circ$, similar to the Sn-end case in Fig.~\ref{fig:theta_dep}(a). 

For comparison, we also studied the bulk case [Fig.~\ref{fig:theta_dep}(c)]. We observe rather smooth and monotonic $\theta$ dependences of $\sigma_{\rm H}$ and $\alpha_{\rm N}$ compared with the monolayers. These behaviors are qualitatively consistent with the previous studies~\cite{ON,Ghimire,OKN}. Also in this bulk case, $\sigma_{\rm H}$ and $\alpha_{\rm N}$ are almost antisymmetric for $\varphi=0^\circ$, but not for $\varphi=90^\circ$~\cite{OKN}. 

Consequently, our results highlight the distinctive transport behaviors while changing the magnetization direction in the monolayer systems. In particular, the Sn-end monolayer exhibits successive discrete changes of $\sigma_{\rm H}$, indicating topological transitions between different anomalous quantum Hall insulators. In addition, this behavior appears differently for $\varphi=0^\circ$ and $90^\circ$, suggesting that the topological transitions take place in a different manner depending on the tilting direction. These findings will be discussed from the viewpoints of the electronic band structure and the Berry curvature in the next section.

\section{Electronic band structure and Berry curvature}
\label{sec:band_top}

\begin{figure}[t]
\includegraphics[width=70mm]{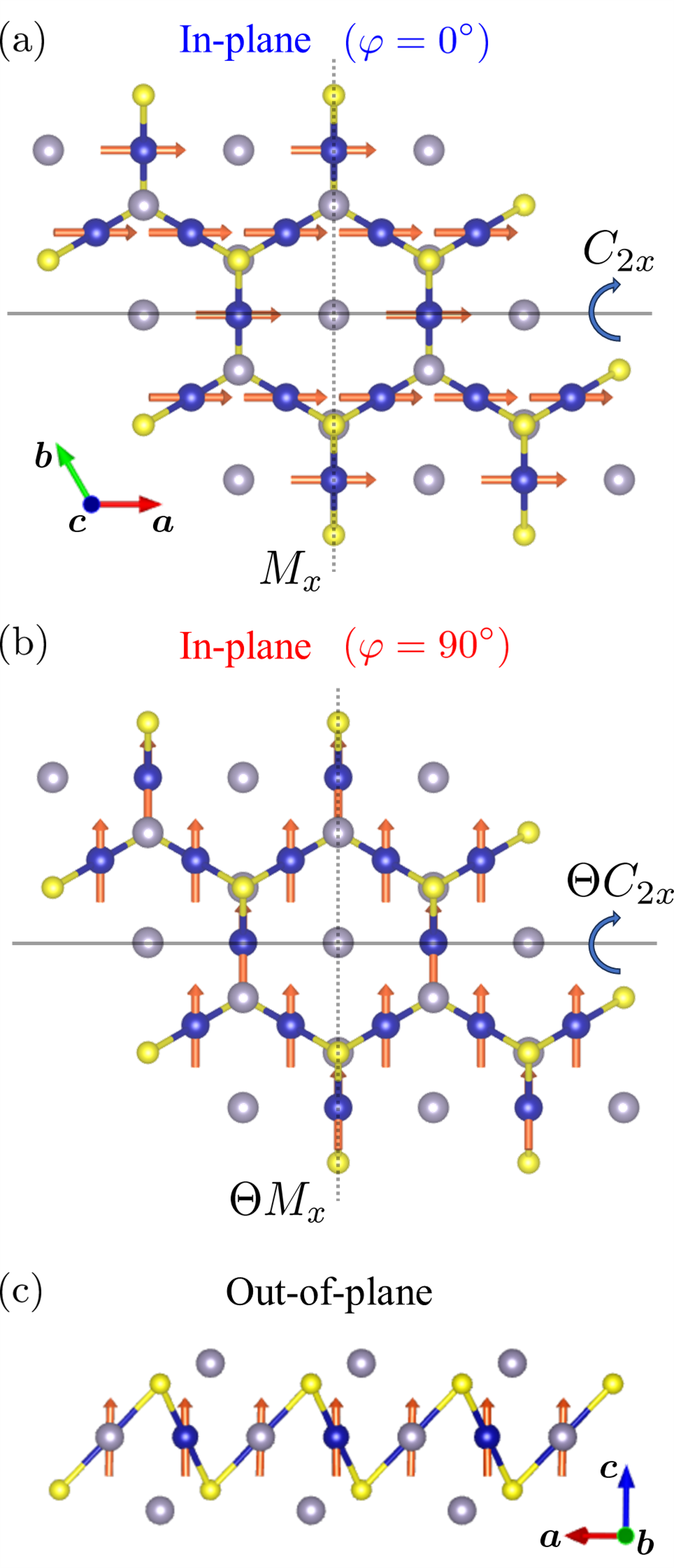}
\caption{Schematic pictures of (a) the in-plane ($\varphi = 0^\circ$), (b) in-plane ($\varphi = 90^\circ$), and (c) out-of-plane directions of the ferromagnetic moments, which are parallel to ${\bm a}$, ${\bm c} \times {\bm a}$, and ${\bm c}$, respectively. The orange arrows represent the Co magnetic moments. The relevant symmetries to our analysis are denoted in (a) and (b).}
\label{fig:symmetry}
\end{figure}

In this section, we present the electronic band structures obtained by the {\it ab initio} calculations, and discuss the relation between their angular dependences and the resuts in the previous section. Figure~\ref{fig:band} shows the electronic band structures along the symmetric lines in the Brillouin zones shown in Fig.~\ref{fig:Lattice}(c) near the Fermi level set at zero while changing the tilting angle $\theta$ for (a) the Sn-end monolayer, (b) the S-end monolayer, and (c) the bulk. The upper and lower panels display the results for $\varphi = 0^\circ$ and $\varphi = 90^\circ$, respectively. 

We first discuss the case of Sn-end monolayer [Fig.~\ref{fig:band}(a)]. In this case, the system at $\theta=0^\circ$, namely, with the out-of-plane magnetic moment, is an anomalous quantum Hall insulator with the Chern number $C=-3$ with the Weyl gap opening along the $\Gamma$-K line~\cite{Muechler,NKM}. While tilting the moment by increasing $\theta$, the Weyl gap decreases and closes around $\theta = 90^\circ$ for $\varphi=0^\circ$, as shown in the upper panel of Fig.~\ref{fig:band}(a); see also Appendix~\ref{sec:berry_3d}. In this case, the change of the band structure is almost antisymmetric with respect to $\theta=90^\circ$; for instance, the results for $\theta = 73^\circ$ and $\theta = 108^\circ$ are almost equivalent. Meanwhile, for $\varphi=90^\circ$, the Weyl gap similarly decreases and closes around $\theta = 69^\circ$ as shown in the lower panel of Fig.~\ref{fig:band}(a)~\cite{comm1}. 

In Fig.~\ref{fig:Berry_Sn_mono}, we show the angular dependence of the Berry curvature of the band just below the Fermi level in the Sn-end monolayer. We present the Berry curvature of the other bands near the Fermi level in Appendix~\ref{sec:berry_3d}. In the case of $\varphi = 0^\circ$ [Fig.~\ref{fig:Berry_Sn_mono}(a)], we find that the sign reversal of the Berry curvature associated with the Weyl gap closing occurs (i) between $\theta=76^\circ$ and $\theta=80^\circ$, (ii) at $\theta = 90^\circ$, (iii) between $\theta = 101^\circ$ and $\theta = 105^\circ$. These correspond to the topological transitions inferred by $\sigma_{\rm H}$ in Fig.~\ref{fig:theta_dep}(a) with the changes of the Chern number as (i) $C=-3 \to -1$, (ii) $C=-1 \to +1$ via $C=0$ at $\theta=90^\circ$, and (iii) $C=1 \to 3$. We note that the Berry curvature along the $k_x = 0$ axis remains at $\theta=90^\circ$ since the very small Weyl gap remains due to small spin canting; the gap should be closed if the magnetic moments are perfectly parallel to the $a$ axis; see the symmetry argument in Sec.~\ref{sec:sym}. In contrast, in the case of $\varphi = 90^\circ$ [Fig.~\ref{fig:Berry_Sn_mono}(b)], there are two topological transitions (iv) between $\theta=69^\circ$ and $\theta=79^\circ$ and (v) between $\theta=93^\circ$ and $\theta=99^\circ$, where the Chern number changes as $C=-3 \to -1$ and $-1 \to 3$, respectively. These correspond well to the $\theta$ dependence of $\sigma_{\rm H}$ in Fig.~\ref{fig:theta_dep}(a). The $\theta$ evolution of the Berry curvature is asymmetric with respect to $\theta=90^\circ$, and the number of topological transitions is one less than in $\varphi=0^\circ$, because the two pairs of Weyl gaps close simultaneously at the transition (v). The asymmetric change leaves the nonzero Chern number at $\theta=90^\circ$ (see also Appendix~\ref{sec:berry_3d}), which leads to the planar quantized anomalous Hall effect discussed in Sec.~\ref{sec:transport_fs}. 

Next, we discuss the results for the S-end monolayer [Fig.~\ref{fig:band}(b)]. In this case, the system is always metallic for both $\varphi=0^\circ$ and $90^\circ$, which prevents $\sigma_{\rm H}$ from being quantized as found in Fig.~\ref{fig:theta_dep}(b). Furthermore, there are a greater number of gapped Weyl nodes at low energy than the Sn-end monolayer; see also Appendix~\ref{sec:berry_3d}. These Weyl gaps close and cause topological transitions at various magnetization angles, leading to the complicated $\theta$ dependences in Fig.~\ref{fig:theta_dep}(b). Nevertheless, for $\varphi=0^\circ$, the band structure changes almost antisymmetric with respect to $\theta$, as represented by almost the same band structures for $\theta = 45^\circ$ and $\theta = 135^\circ$. This is consistent with almost antisymmetric behavior of $\sigma_{\rm H}$ and $\alpha_{\rm N}$ in Fig.~\ref{fig:theta_dep}(b), similar to the Sn-end case. 

Finally, we briefly discuss the bulk result [Fig.~\ref{fig:band}(c)]. The system is always metallic, similar to the S-end monolayer case. Besides, the band structure around the Fermi level is much simpler and the number of Weyl gaps are more scarce than the S-end monolayer, leading to the simpler $\theta$ dependences of $\sigma_{\rm H}$ and $\alpha_{\rm N}$. We observe the gap closing on the M$'$-$\Gamma$ line at $\theta=90^\circ$ and the $\Gamma$-$\rm K'$ line at $\theta=135^\circ$ in the $\varphi=0^\circ$ case, and the M$'$-$\Gamma$ line at $\theta=132^\circ$ in the $\varphi=90^\circ$ case. 

\section{Symmetry argument}
\label{sec:sym}

\begin{figure*}
\hspace*{-5mm}
\includegraphics[width=160mm]{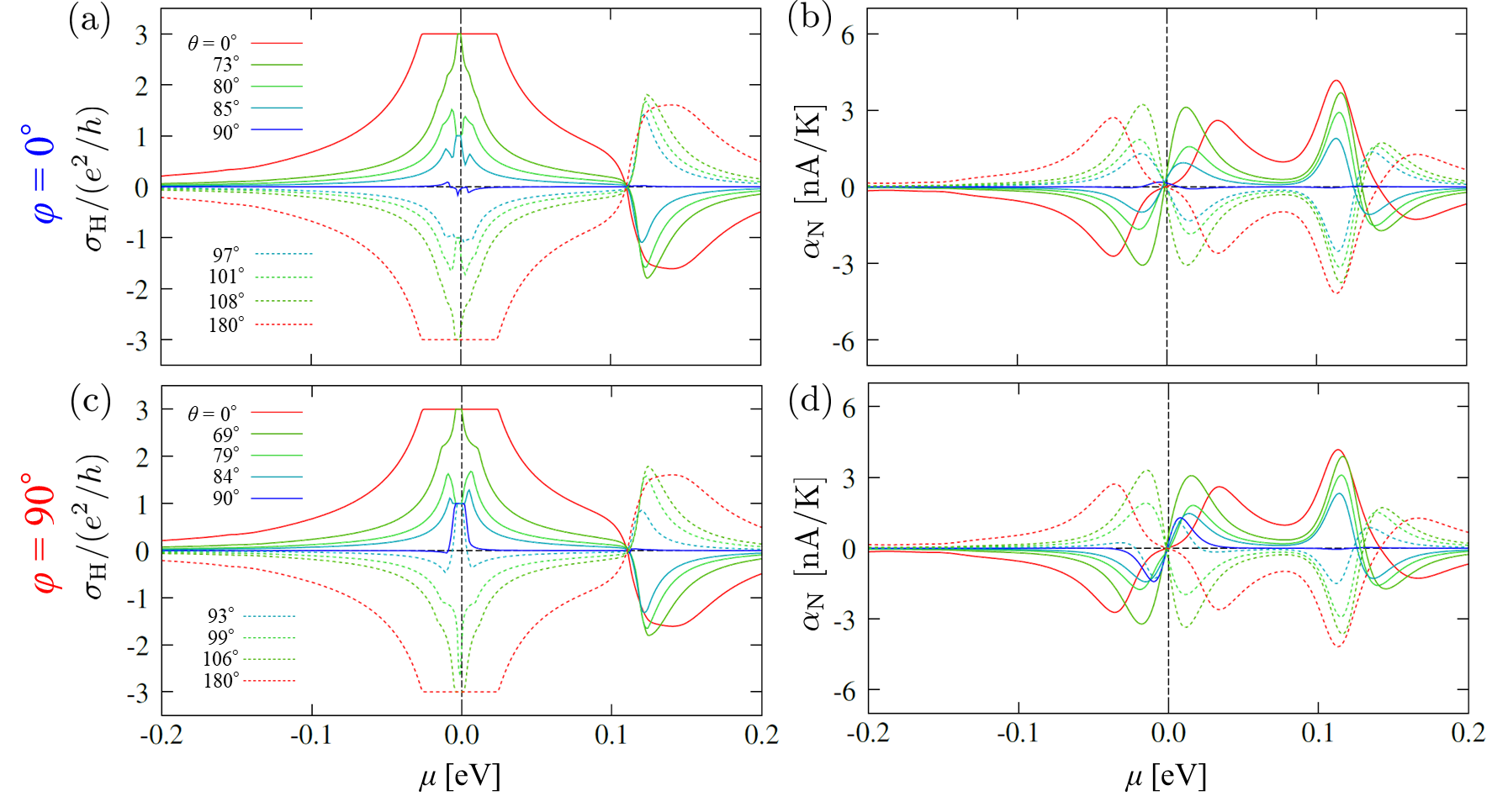}
\caption{Chemical potential dependences of [(a) and (c)] the anomalous Hall conductivity $\sigma_{\rm H}$ at zero temperature and [(b) and (d)] the anomalous Nernst conductivity $\alpha_{\rm N}$ at the temperature of $5$~meV in the Sn-end monolayer. The data are calculated for $\varphi = 0^\circ$ in (a) and (b), and for $\varphi = 90^\circ$ in (c) and (d), at several tilting angles $\theta$.}
\label{fig:ahe}
\end{figure*}

\begin{figure*}
\includegraphics[width=175mm]{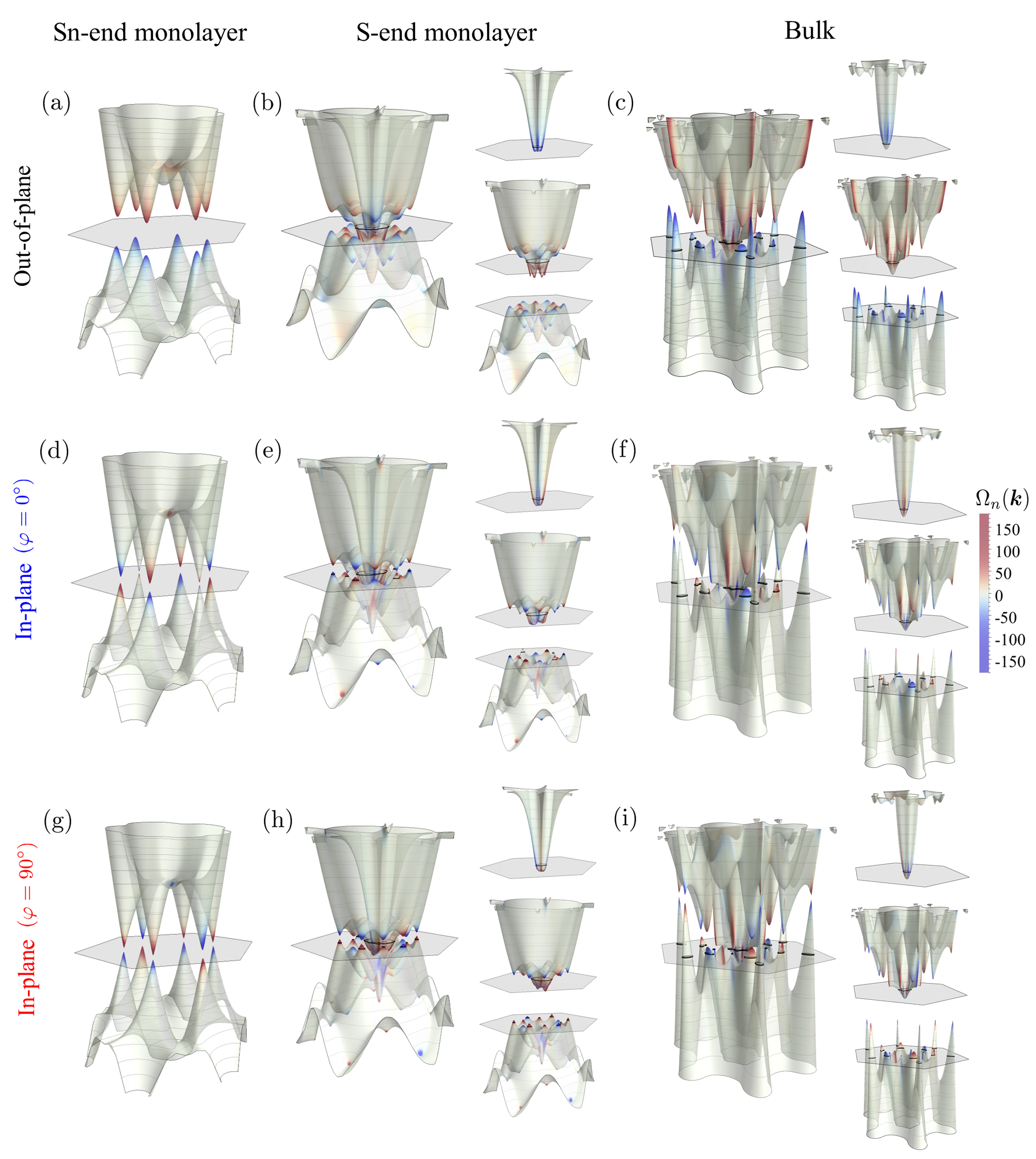}
\caption{Berry curvature with the band structure of [(a), (d), (g)] Sn-end monolayer, [(b), (e), (h)] S-end monolayer, and [(c), (f), (i)] $k_z = 0$ cut of bulk for the [(a)--(c)] out-of-plane, [(d)--(f)] in-plane ($\varphi=0^\circ$), and [(g)--(i)] in-plane ($\varphi=90^\circ$) magnetic moments in the energy range from $-0.2$~eV to $0.2$~eV. The black lines on the gray hexagons in each panel represent the Fermi surfaces. The small panels in the row of S-end monolayer and bulk present the results in each band separately.}
\label{fig:Berry_Sn}
\end{figure*}

Let us discuss the band topology from the symmetry point of view. As mentioned in Sec.~\ref{sec:methods}, the lattice structures of the monolayers are symmetrized after the optimization, which have threefold rotational symmetry around the $c$ axis, $C_{3c}$, twofold rotational symmetry around the $a$ axis, $C_{2x}$, and mirror symmetry perpendicular to the $a$ axis, $M_x$. 

We first focus on the $\varphi=0^\circ$ case. When $\theta = 90^\circ$ where all the magnetic moments are perfectly oriented along the ${\bm a}$ axis [Fig.~\ref{fig:symmetry}(a)], all the above lattice symmetries are preserved. Hence, in this case, the Berry curvature should satisfy the relations 
\begin{align}
\Omega(k_x, -k_y)|_{\theta=90^\circ, \varphi=0^\circ} = -\Omega(k_x, k_y)|_{\theta=90^\circ, \varphi=0^\circ}, \\ 
\Omega(-k_x, k_y)|_{\theta=90^\circ, \varphi=0^\circ} = -\Omega(k_x, k_y)|_{\theta=90^\circ, \varphi=0^\circ}. 
\end{align}
Note that the band index $n$ for the Berry curvature is dropped here and hereafter. These lead to the cancellation of the Berry curvature in each band, and consequently, the disappearance of the anomalous Hall and Nernst conductivities. In particular, these relations indicate $\Omega(k_x, 0) = \Omega(0, k_y) = 0$, meaning that the Berry curvature vanishes on the $k_x$ and $k_y$ axes. These arguments are indeed approximately satisfied in the {\it ab initio} results in the previous sections. 

We can extend the argument to general $\theta$. Since the state with $\theta = 90^\circ - \gamma$ and $\theta = 90^\circ + \gamma$ are related with each other via the $C_{2x}$ or $M_x$ operation, we obtain the relations
\begin{align}
\Omega(k_x, -k_y)|_{\theta=90^\circ - \gamma, \varphi=0^\circ} = -\Omega(k_x, k_y)|_{\theta=90^\circ + \gamma, \varphi=0^\circ}, \label{eq:O_x1} \\ 
\Omega(-k_x, k_y)|_{\theta=90^\circ - \gamma, \varphi=0^\circ} = -\Omega(k_x, k_y)|_{\theta=90^\circ + \gamma, \varphi=0^\circ}. \label{eq:O_x2}
\end{align}
These are consistent with the results in Fig.~\ref{fig:Berry_Sn_mono}(a). Equations~\eqref{eq:O_x1} and \eqref{eq:O_x2} lead to the relations of $\sigma_{\rm H}$ and $\alpha_{\rm N}$ as 
\begin{align}
\sigma_{\rm H}|_{\theta=90^\circ - \gamma, \varphi=0^\circ} = -\sigma_{\rm H}|_{\theta=90^\circ + \gamma, \varphi=0^\circ}, \label{eq:sym_x} \\
\alpha_{\rm N}|_{\theta=90^\circ - \gamma, \varphi=0^\circ} = -\alpha_{\rm N}|_{\theta=90^\circ + \gamma, \varphi=0^\circ}, \label{eq:sym_x2}
\end{align}
respectively. These explain well our {\it ab initio} results in Fig.~\ref{fig:theta_dep}. 

Next we consider the case of $\varphi=90^\circ$. In Fig.~\ref{fig:symmetry}(b), we illustrate the situation of in-plane ferromagnetic state with $\theta = 90^\circ$ as a typical example. In this case, the system remains invariant under the $\Theta C_{2x}$ and $\Theta M_x$ operations, where $\Theta$ is a time-reversal operator. This is because the magnetic moment is parallel to the mirror plane. Since the same argument can be made for general $\theta$, the Berry curvature should satisfy the relations 
\begin{align} 
\Omega(k_x, -k_y)|_{\varphi = 90^\circ} = \Omega(k_x, k_y)|_{\varphi = 90^\circ}, \label{eq:O_y1} \\
\Omega(-k_x, k_y)|_{\varphi = 90^\circ} = \Omega(k_x, k_y)|_{\varphi = 90^\circ}, \label{eq:O_y2}
\end{align} 
for all $\theta$. 
These are consistent with the {\it ab initio}-based analysis shown in Fig.~\ref{fig:Berry_Sn_mono}(b) and allow the appearance of the anomalous Hall and Nernst effects~\cite{Ghimire}, including the planar quantized anomalous Hall effect found in Fig.~\ref{fig:theta_dep}(a).  

From the above discussion, we can predict the behavior of transport properties when the magnetization is rotated in the kagome plane ($\theta = 90^\circ$). Since $\sigma_{\rm H}=0$ when $\varphi=0^\circ$, it should vanish when $\varphi=(60n)^\circ$ ($n$ is an integer). Similarly, $\sigma_{\rm H}/(e^2/h)=\pm 1$ when $\varphi = (\mp 30+120n)^\circ$. Thus, when the magnetization is rotated in the kagome plane, $\sigma_{\rm H}$ changes as $0 \to +1 \to 0 \to -1 \to 0 \cdots$ in unit of $e^2/h$ with $120^\circ$ period. 

Finally, we comment on the case of out-of-plane moment [Fig.~\ref{fig:symmetry}(c)]. In this case, Eqs.~(\ref{eq:O_x1}) and (\ref{eq:O_x2}) with $\gamma=90^\circ$ give the relations between the two cases of out-of-plane moments, $\theta = 0^\circ$ and $\theta=180^\circ$. In addition, Eqs.~(\ref{eq:O_y1}) and (\ref{eq:O_y2}) hold. These relations are again consistent with the results in the previous sections. 

\section{Summary}
\label{sec:summary}

To summarize, focusing on the thin films of Co-based shandite with Co kagome monolayer, we have investigated the magnetization angle dependences of the electronic state, band topology, and transport properties. Using the {\it ab initio} calculations, we revealed distinctive behaviors depending on the surface termination of the films. In the Sn-end monolayer, we found successive topological phase transitions between the Chern insulating states with different Chern numbers while rotating the magnetization direction, resulting in discrete changes in the anomalous Hall conductivity $\sigma_{\rm H}$ between the corresponding quantized values. Remarkably, we discovered the planar quantized anomalous Hall effect: $\sigma_{\rm H}$ is quantized with nonzero Chern number even when the magnetization is oriented in the in-plane direction perpendicular to a Co-Co bond. In contrast, in the S-end monolayer, $\sigma_{\rm H}$ as well as the anomalous Nernst conductivity $\alpha_{\rm N}$ shows moderate changes without quantization as the system is always metallic. We show the origins of these behaviors by careful analyses of the Berry curvature distributions in each band and the symmetry arguments. Our results pave the way for controlling topological properties of the Weyl semimetal toward device applications, by leveraging various tuning methods for the magnetic anisotropy in thin films, such as magnetostriction, substrate effects, and thickness changes. 

\section*{Acknowledgements} 
The authors thank K. Fujiwara, K. Kobayashi, K. Nomura, S. Okumura, A. Ozawa, and A. Tsukazaki for fruitful discussions. 
This work is supported by JST CREST (Grant No. JPMJCR18T2) and JSPS KAKENHI (Grant No. JP21K13875 and JP22K03509). 

\appendix

\section{Chemical potential dependences of transport properties in the Sn-end monolayer}
\label{sec:transport}

In this Appendix, we present the chemical potential dependences of the anomalous Hall conductivity $\sigma_{\rm H}$ and the anomalous Nernst conductivity $\alpha_{\rm N}$ for the Sn-end monolayer. Figure~\ref{fig:ahe} show the results calculated for the band structures at several values of $\theta$, by changing the chemical potential in Eqs.~\eqref{eq:AHE} and \eqref{eq:ANE}. As discussed in Secs.~\ref{sec:transport_fs} and \ref{sec:band_top}, the Sn-end monolayer is in the Chern insulating states with different quantized values of $\sigma_{\rm H}$ depending on the moment direction. Correspondingly, we find integer plateaus around zero chemical potential: $\sigma_{\rm H}/(e^2/h)=+3$, $+1$, $-1$, and $-3$ for $\varphi=0^\circ$ [Fig.~\ref{fig:ahe}(a)], and $\sigma_{\rm H}/(e^2/h)=+3$, $+1$, and $-3$ for $\varphi=90^\circ$ [Fig.~\ref{fig:ahe}(c)]. These plateaus represent the Weyl gaps which make the system being the Chern insulators. In addition, in the case of $\varphi=0^\circ$, weconfirm that the relations in Eqs.~(\ref{eq:sym_x}) and \eqref{eq:sym_x2} are approximately satisfied at all values of the chemical potential. We also note that $\alpha_{\rm N}$ shown in Figs.~\ref{fig:ahe}(b) and \ref{fig:ahe}(d) are approximately given by the chemical potential derivative of $\sigma_{\rm H}$, as expected from the generalized Mott relation~\cite{XYFN}. 

\section{Berry curvature in each band}
\label{sec:berry_3d}

In this Appendix, we show the Berry curvature in each band for Sn-end and S-end monolayers as well as the bulk. Figures~\ref{fig:Berry_Sn}(a), \ref{fig:Berry_Sn}(d), and \ref{fig:Berry_Sn}(g) show that the Sn-end monolayer is in the distinct Chern insulating states for the out-of-plane, in-plane $(\varphi = 0^\circ)$, and in-plane $(\varphi = 90^\circ)$ moments, respectively. We can also observe that the gap is almost closed on the $k_x$ and $k_y$ axes in the in-plane $x$ case from Fig.~\ref{fig:Berry_Sn}(d). Note that the Berry curvatures at $\theta = 90^\circ$ shown in Fig.~\ref{fig:Berry_Sn_mono} correspond to the results in the lower band depicted in Figs.~\ref{fig:Berry_Sn}(d) and \ref{fig:Berry_Sn}(g). In the S-end cases shown in Figs.~\ref{fig:Berry_Sn}(b), \ref{fig:Berry_Sn}(e), and \ref{fig:Berry_Sn}(h), the distribution of the Weyl nodes where the Berry curvature becomes intense is complicated as discussed in Secs.~\ref{sec:transport_fs} and \ref{sec:band_top}. Figures~\ref{fig:Berry_Sn}(c), \ref{fig:Berry_Sn}(f), and \ref{fig:Berry_Sn}(i) are the similar plots for the bulk in the $k_z = 0$ cut, again showing the consistency with the symmetry arguments.

\end{document}